\documentclass[fleqn,10pt]{wlscirep}

\usepackage{graphicx}
\usepackage{amsmath}
\usepackage{verbatim} 
\usepackage{color}
\usepackage{url}
\usepackage{cite}

\title{High-$Q$ photonic resonators and electro-optic coupling using silicon-on-lithium-niobate}

\author[1,*]{Jeremy D. Witmer}
\author[1]{Joseph A. Valery}
\author[1]{Patricio Arrangoiz-Arriola}
\author[1]{Christopher J. Sarabalis}
\author[1]{Jeff T. Hill}
\author[1,+]{Amir H. Safavi-Naeini}
\affil[1]{Ginzton Laboratory, Stanford University, Stanford, California 94305, USA}
\affil[*]{jwitmer@stanford.edu}
\affil[+]{safavi@stanford.edu}

\begin{abstract}
Future quantum networks in which superconducting quantum processors are connected via optical links,  will require microwave-to-optical photon converters that preserve entanglement.  A doubly-resonant electro-optic modulator (EOM) is a promising platform to realize this conversion.   Here, we present our progress towards building such a modulator by demonstrating the optically-resonant half of the device.  We demonstrate high quality ($Q$) factor ring, disk and photonic crystal resonators using a hybrid silicon-on-lithium-niobate material system.  Optical $Q$ factors up to 730,000 are achieved, corresponding to propagation loss of 0.8 dB/cm.  We also use the electro-optic effect to modulate the resonance frequency of a photonic crystal cavity, achieving a electro-optic modulation coefficient between 1 and 2 pm/V.  In addition to quantum technology, we expect that our results will be useful both in traditional silicon photonics applications and in high-sensitivity acousto-optic devices.

\end{abstract}

\begin{document}

\flushbottom
\maketitle

\section*{Introduction}

Electro-optic modulators (EOMs) are an integral part of modern communication, enabling the long-haul fiber optic links which form the backbone of the internet.  In order to realize a quantum internet~\cite{Kimble2008} in which superconducting quantum processors~\cite{Devoret2013} are connected via optical links, a quantum-efficient electro-optic modulator is required to convert signals between the microwave and optical domains while preserving quantum correlations. Optically and electrically resonant electro-optic modulators can be operated as such quantum frequency converters. With sufficiently high electro-optic coupling and $Q$ factors, these converters may be used to convert a single microwave photon to a single optical photon and vice versa~\cite{Tsang2010, Tsang2011} in a process analogous to quantum frequency conversion between light fields mediated by optical~\cite{Kumar1990} or optomechanical\cite{Safavi-Naeini2010a,Hill2012} nonlinearities. A variety of approaches have been pursued to implement microwave-to-optical conversion, some of which are based on enhancement of the effective electro-optic response by inclusion of an intermediate element, e.g. mechanical~\cite{Bochmann2013a,Andrews2014,Balram2015b,Dieterle2016} or magnonic~\cite{Tabuchi2015,Bourhill2016,Hisatomi2016} resonances.  A more direct approach is to use an optical resonator that is modulated by a microwave signal via the linear electro-optic effect in a material with a sufficiently large electro-optic coefficient. Lithium niobate (LN, LiNbO$_3$) is an exemplary electro-optic material, and recently centimeter scale lithium niobate optical resonators embedded in 3D microwave cavities have been pursued in this context~\cite{Rueda2016}. Our approach utilizes LN and is focused on developing a platform where the optical and microwave resonators are co-integrated on a single chip.

Here, we present progress towards an on-chip, integrated quantum electro-optic converter based on the hybrid silicon-on-lithium niobate (SiLN) platform.  We demonstrate high-$Q$ ring and disk resonators as well as the first photonic crystal resonators in the SiLN platform, and measure the electro-optic coefficient for a photonic crystal cavity.  After briefly describing the fabrication process and experimental setup, we present optical measurements of these resonators showing large measured $Q$ factors (summarized in Table \ref{tab:Qs}).  Finally, we demonstrate electro-optic tuning of the photonic crystal cavity resonance, achieving a maximum modulation coefficient of 1.93 pm/V.

\section*{Material Considerations}

Making an on-chip, integrated quantum EOM that operates at low temperatures while achieving long optical and microwave coherence times is challenging.  Since high-$Q$ factor resonators are required for both microwaves and light, it is important to use materials with low loss at both microwave and optical frequencies. Microwave loss in materials at low temperatures and at low drive powers make the material requirements distinct from classical modulators. In particular, two-level system (TLS) losses  at microwave frequencies\cite{Pappas2011} prevent the use of polymers and amorphous materials. This rules out using silicon-organic hybrid (SOH) platforms that allow extremely efficient modulation~\cite{Leuthold2009, Leuthold2013}, as well as adhesive layers such as divinylsiloxane-bis-benzocyclobutene (BCB) which has been used for silicon/III-V~\cite{Roelkens2010} and silicon/LN heterogeneous integration\cite{Guarino2007}.

Our approach is to fabricate a heterogeneously integrated system that only uses materials known to have low optical and microwave losses. One such material, lithium niobate, is an artificial ferroelectric  with a non-centrosymmetric trigonal crystal structure.  Because of its large piezoelectric, photo-elastic, $\chi^{(2)}$, and electro-optic (EO) coefficients \cite{Weis1985}, LN has become ubiquitous in nonlinear optics, acoustic wave filters, acousto-optic devices,  and electro-optic modulators.  LN has also been shown to support microwave modes with $Q$'s on the order of $10^5$ at millikelvin temperatures and single photon power levels\cite{Goryachev2015}.  However, the fabrication of high-$Q$ optical resonators in LN remains challenging.  To date the main strategies for patterning optical nano-structures into LN have been femtosecond laser ablation \cite{Lin2015a}, focused ion beam (FIB) milling \cite{Lin2015a, Diziain2015, Dahdah2011}, argon ion milling \cite{Wang2014,Wang2013}, or wet etching of proton-exchanged regions \cite{Chen1995}.  These techniques typically lead to limited minimum feature sizes as well as large sidewall roughness which causes unwanted scattering of light and reduces optical $Q$'s at smaller mode volumes where surface effects begin to dominate.  In addition, FIB milling is an inherently time-intensive process which is not well-suited to large-scale integration.  We note that there are other electro-optic materials, such as gallium arsenide, in which photonic crystals can also be patterned directly \cite{Buckley2014a} - however, none of these direct patterning approaches have achieved $Q$'s as high as those reported in silicon photonic crystal resonators.

Hybrid material systems provide a way to take advantage of the excellent properties of active materials, such as LN, while simultaneously leveraging the powerful, scalable fabrication processes associated with silicon. We focus particularly on silicon-on-lithium-niobate (SiLN), in which single-crystal silicon photonic structures are directly bonded onto an LN substrate \cite{Chiles2014,Weigel2015,Witmer2016}.  In SiLN devices, the optical mode is confined by the silicon layer while still retaining modal overlap with the LN substrate. In these systems, about 20\% of the optical field energy can be in the lithium niobate allowing the mode to be highly confined by the subwavelength patterning of the silicon film while still retaining a large electro-optic coefficient.

\section*{Fabrication and Methods}

\begin{figure}
    \centering
\includegraphics[width=5in]{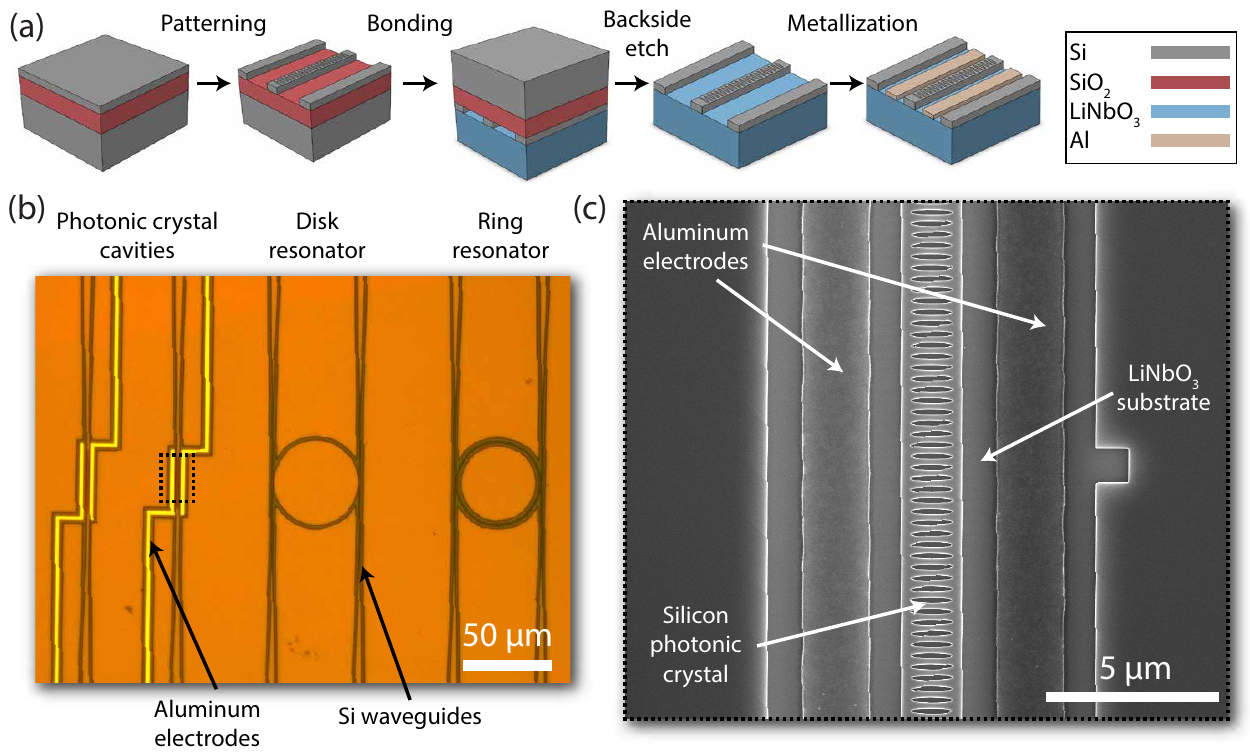}
    \caption{(a) The SiLN fabrication process, showing e-beam lithography and etching of photonic structures into an SOI die, bonding  to LN, removing the backside silicon, and patterning electrodes.  (b) An optical microscope image showing fabricated ring, disk and photonic crystal structures.  (c) A scanning electron micrograph image of a SiLN photonic crystal with electrodes.}
    \label{fig:fab_steps}
\end{figure}

The fabrication flow for the hybrid silicon/lithium niobate structures is similar to the processes described in references [\citenum{Chiles2014, Witmer2016, Weigel2015 }] and is illustrated in Figure \ref{fig:fab_steps}(a).  The process starts with a silicon-on-insulator (SOI) die with a 145 nm top silicon layer.  Photonic structures are patterned onto the SOI die using electron beam lithography and an ICP-RIE etch tool (Lam 9400 TCP Poly Etcher using Cl$_2$ and HBr gas chemistry).  Next, the SOI die and an X-cut LN die are cleaned using standard methods such as solvent sonication and piranha (H$_2$SO$_4$ and H$_2$O$_2$).  The die surfaces are then activated via exposure to an O$_2$ plasma, a process which creates dangling bonds and extremely hydrophilic surfaces, and then manually bonded together at room temperature.  After bonding, an SF$_6$ RIE etch is used to remove the bulk silicon from the SOI die and a buffered oxide etch strips the buried oxide.  Finally, for electro-optic devices, a second layer of electron beam lithography is used to pattern electrodes, followed by aluminum evaporation and lift-off.

\begin{figure}
    \centering
\includegraphics[width=5in]{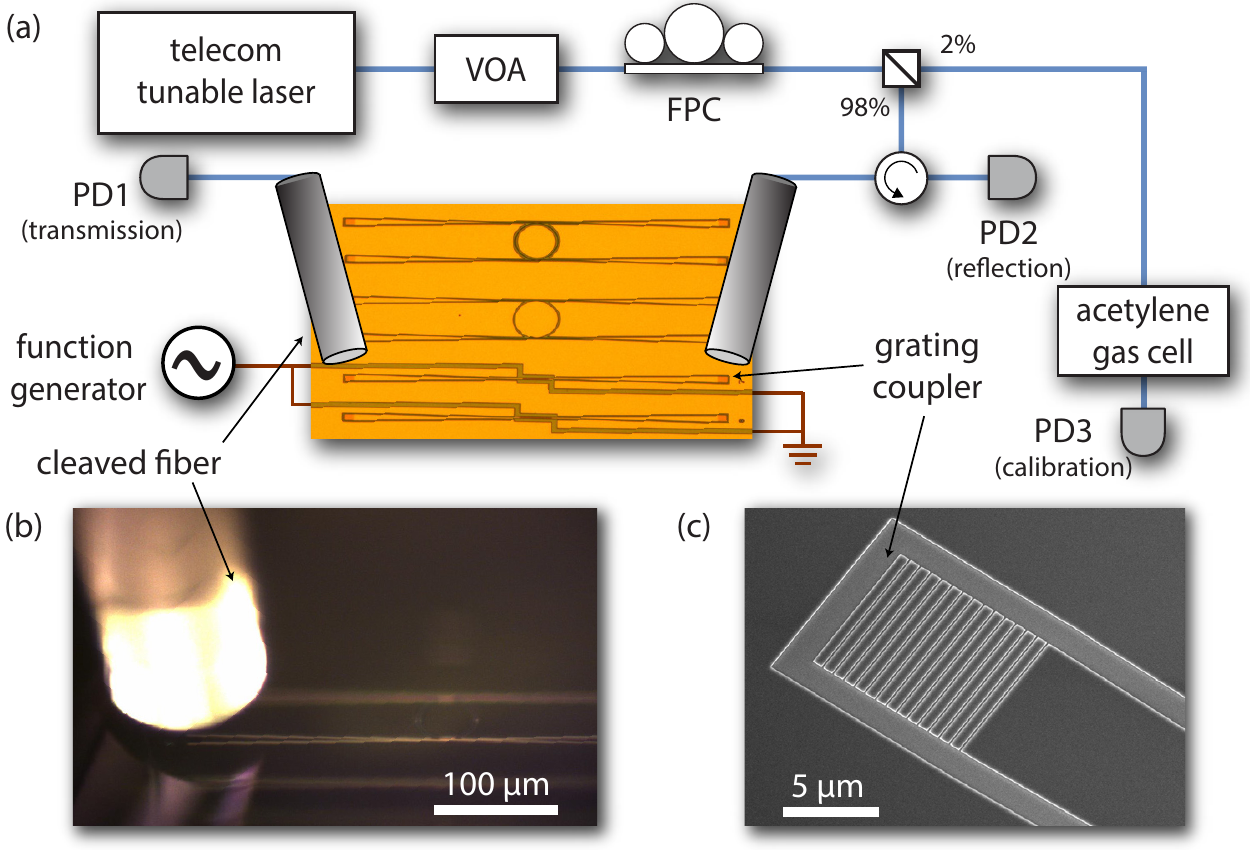}
    \caption{(a) Diagram of the experimental setup.  (b) A microscope image showing a cleaved fiber coupling light to a ring resonator.  (c) Scanning electron micrograph of a typical SiLN grating coupler used in these experiments. }
    \label{fig:setup}
\end{figure}

The experimental set-up used to characterize the SiLN photonic structure is shown in Figure \ref{fig:setup}.  A Santec tunable-wavelength telecom laser source (Santec TSL-550-A) is used to probe the devices optically, and the reflected and transmitted power is recorded using Newport photodetectors.  The power level and polarization are controlled using a variable optical attenuator (VOA) and fiber polarization controller (FPC), respectively. In order to couple light in and out of the SiLN waveguides, cleaved single-mode optical fibers, oriented at 15 degrees from vertical, are aligned to on-chip grating couplers (Figure \ref{fig:setup} (c)).  The grating couplers used here have a transmission window which is approximately Gaussian, with a FWHM bandwidth of $\approx$54 nm and a measured peak efficiency of 8\%.  The grating coupler design efficiency is limited by the high index of LN ($\approx$~2.2) and the relatively thin (145 nm) silicon layer; however, with improved designs we have observed efficiencies of 16\%.  In addition, about 2\% of the light from the laser source was split off and sent through an acetylene reference cell to another detector.  The acetylene absorption lines were fit to Lorentzian functions during post-processing and used to provide an independent calibration of the laser wavelength scan (accurate to within 2-3 pm).

\section*{Results and Discussion}

\subsection*{Ring Resonators}

\begin{figure}[ht]
\begin{center}
\includegraphics[width=5in]{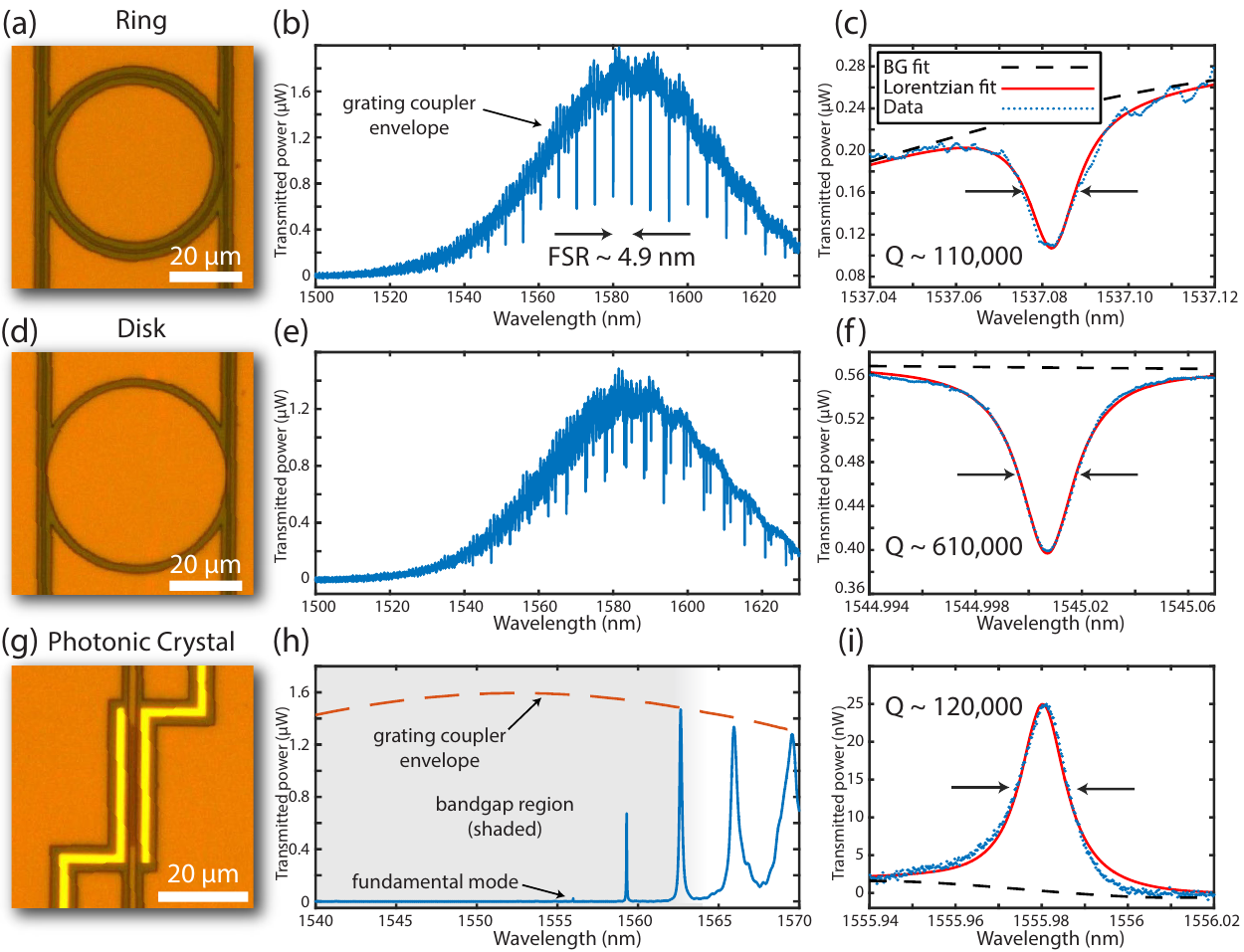}
\caption{(a,d,g) Optical microscope images showing a microring, microdisk and photonic crystal respectively.  For each structure, a spectrum of transmitted optical power vs. wavelength is shown (b,e,h).  In (b) and (e) the Gaussian transmission window of the grating couplers is apparent, while in (h) the dashed-red line shows approximately the grating coupler envelope.  The enlarged plots (c,f,i) show high-$Q$ modes for each of the three structures.  Loaded $Q$'s were calculated by fitting a Lorentzian function to the peaks/dips.  The dashed black line shows a polynomial fit to the background. }
\label{fig:scans}
\end{center}
\end{figure}

Ring resonators were fabricated as an initial demonstration of resonators in the SiLN platform.  Optical transmission spectra from a typical ring resonator, as well as a disk and photonic crystal cavity, are shown in Figure \ref{fig:scans}.  The ring resonators are side-coupled to two symmetric feed waveguides, one on either side.  As a result, the optical modes appear as dips in the transmission (Figure \ref{fig:scans}(b)).  The fast oscillations superimposed on the grating coupler transmission window are due to reflections between grating couplers on either end of the feed waveguide. 

The rings have a diameter of 46 $\mu$m and a measured free spectral range (FSR) of 4.9 nm.  The example mode in Figure \ref{fig:scans} (c) is undercoupled, with a loaded $Q$ of approximately 110,000. The intrinsic optical quality factor is estimated from the transmission contrast to be approximately 160,000.  At 1550 nm, the bending loss limited $Q$ is simulated to greatly exceed 1 million, suggesting that this is not a critical loss channel for these devices.  

\begin{table}[b]
\centering
\caption{Measured loaded and unloaded quality factors for SiLN photonic structures.  The loaded $Q$'s are measured by fitting Lorentzians to the resonances, while the intrinsic $Q$'s are estimates inferred from the transmission contrast.  All modes are undercoupled.}
\begin{tabular}{|c|c|c|}
\hline
\textbf{Structure} & \textbf{Loaded $Q$} & \textbf{Intrinsic $Q$} \\ 
\hline
Ring & 110,000 & 160,000 \\
Disk & 610,000 &  730,000 \\
Photonic crystal cavity & 120,000 &  140,000 \\
\hline
\end{tabular}
\label{tab:Qs}
\end{table}

\subsection*{Disk Resonators}
 Disk resonators were fabricated as a second test structure.  Since the higher order modes of disk resonators are further from the silicon side walls compared to the rings, the surface effects on the optical modes should be reduced and the optical quality factors are expected to be higher.  The transmission spectrum of a disk in Figure \ref{fig:scans} (e) shows three distinct families of optical modes which have approximately the same FSR but different coupling strengths, as expected.  For the more weakly coupled modes (which lie closer to the interior of the disk), loaded optical $Q$'s of up to 610,000 were observed (Figure \ref{fig:scans} (f)).  These modes have an FSR of 5.0 nm and are undercoupled.  The loaded $Q$ of 610,000 presented here should be compared to the results of Chen et al., in which they achieved a $Q$ of 14,000 in a hybrid Si/LN platform \cite{Chen2014}.  It should also be compared to undercut disks made by etching thin-film LN directly, such as in references \citenum{Wang2013} ($Q = $ 484,000), \citenum{Wang2014} ($Q = $ 102,000) and \citenum{Lin2015b} ($Q = 2.45 \times 10^6$).

Using the quality factors of the disk resonators, it is possible to estimate an upper bound for the intrinsic propagation loss of the SiLN platform.  Since for the disk we have $\text{FSR} \approx \frac{\lambda_0^2}{n_g L}$ and $Q = \frac{2\pi n_g}{\alpha \lambda_0}$, the optical loss coefficient $\alpha$ can be calculated as
\begin{equation}
    \alpha \approx \frac{2 \pi \lambda_0}{Q L ~\text{FSR}}
\end{equation}
where $Q$ is the quality factor and $L$ is the round-trip length around the disk circumference.  The measured intrinsic quality factor of 730,000 implies an intrinsic optical propagation loss of 0.8 dB/cm. These results underscore the potential of the SiLN platform for making high-$Q$ resonators on an electro-optic substrate in addition to the potential of the platform for nonlinear optics \cite{Chang2016}.

\subsection*{Photonic Crystal Cavities}

Quasi-one-dimensional photonic crystal cavities can simultaneously achieve high $Q$'s and sub-wavelength mode volumes \cite{Deotare2009a, Hill2012}. However, fabricating high-$Q$ photonic crystal cavities directly in LN has proven challenging and $Q$'s demonstrated to date are less than $10^3$ \cite{Diziain2015}.  Here we measure $Q's$ exceeding $10^5$ in a SiLN photonic crystal cavity.

The cavity presented here was simulated using finite-element software (COMSOL) and follows a similar design recipe to the one described in our previous work (reference \citenum{Witmer2016}).  The photonic crystal cavity consists of a silicon waveguide patterned with an array of elliptical holes.  These holes create a periodic modulation in the effective mode index, causing a photonic bandgap to open.  A defect is generated by quadratically increasing the hole width towards the center of crystal (by a total of 15\%), while the lattice spacing is kept constant throughout. This defect creates a localized TE-like mode in the center of the cavity.  The dimensions of the photonic crystal cavity are summarized in Table \ref{tab:dimensions}.

\begin{table}[b]
\centering
\caption{Photonic crystal cavity nominal dimensions }
\begin{tabular}{|c|c|}
\hline
 \textbf{Parameter} &  \textbf{Value} \\ 
\hline
Waveguide width & 1800 nm \\
Silicon thickness & 145 nm \\
Hole width (mirror region) & 1040 nm \\
Hole height & 140 nm \\
Lattice spacing & 320 nm \\
Number of holes in defect region & 91 \\
Number of holes in each mirror region & 20 \\
Scaling of lateral dimensions for measured device & $+4\%$ \\
\hline
\end{tabular}
\label{tab:dimensions}
\end{table}

The coupling rate to the feed waveguide, which extends out from the cavity in both directions, can be adjusted by changing the number of nominal ``mirror'' holes on either side of the defect region.  With 20 mirror holes, the simulated extrinsic quality factors (representing coupling to the waveguide) is 230,000 while the simulated intrinsic quality factor is about 850,000.  This cavity is also designed to have significant overlap of the optical field with the substrate, having 19\% of the electromagnetic energy contained within the LN.  The volume of the fundamental mode (defined as $V = \frac{\int \epsilon |\vec{E}|^2 d^3r}{\text{max}\left( \epsilon |\vec{E}|^2\right)}$ \cite{Vuckovic2002}) is approximately 11.2 $(\lambda_0 /n_{Si})^3$.

The transmission spectrum of a photonic crystal cavity is shown in Figure \ref{fig:scans} (h).  Below 1563 nm there is a bandgap region with virtually no transmitted power outside of the resonant modes.  Several transmission peaks can be seen, corresponding to the fundamental mode at approximately 1556 nm as well as several  higher order modes with increasingly longer wavelength and lower $Q$.  Since we are probing the transmission through the photonic crystal cavity, the device is reflective and there is no transmission away from optical resonances, while on resonance optical modes appear as peaks in the spectrum.  The enlarged spectrum (Figure \ref{fig:scans} (i)) shows the fundamental mode of a photonic crystal resonator with a loaded $Q$ of 120,000.

\subsection*{Electro-Optic Coupling}

Having demonstrated high-$Q$ photonic resonances on SiLN, we now turn to measuring the electro-optic effect. Electro-optic modulation can be obtained by patterning electrodes on the two sides of the SiLN photonic crystals so that we can apply an electric field through the LN substrate.  This field $E$ changes the refractive index $n$ of the LN via the linear electro-optic (Pockels) effect, 
\begin{equation}
    \Delta n_{ij} \approx - \frac{1}{2} \sum_k n_{ij}^3 \mathbf{r}_{ijk}  E_k ,
\end{equation}
where $\mathbf{r}$ is the third-rank electro-optic tensor, thereby shifting the optical resonant frequency of the cavity \cite{Witmer2016}.  In the case of LN, the dominant electro-optic component is $r_{33} = 31$ pm/V, which connects an applied electric field along the extraordinary ($Z$) axis to a change in the extraordinary index $n_e$.  To take advantage of this large component, both the optical $E$-field and external $E$-field are aligned to the extraordinary axis of the X-cut LN.

\begin{figure}[ht]
\begin{center}
\includegraphics[width=5in]{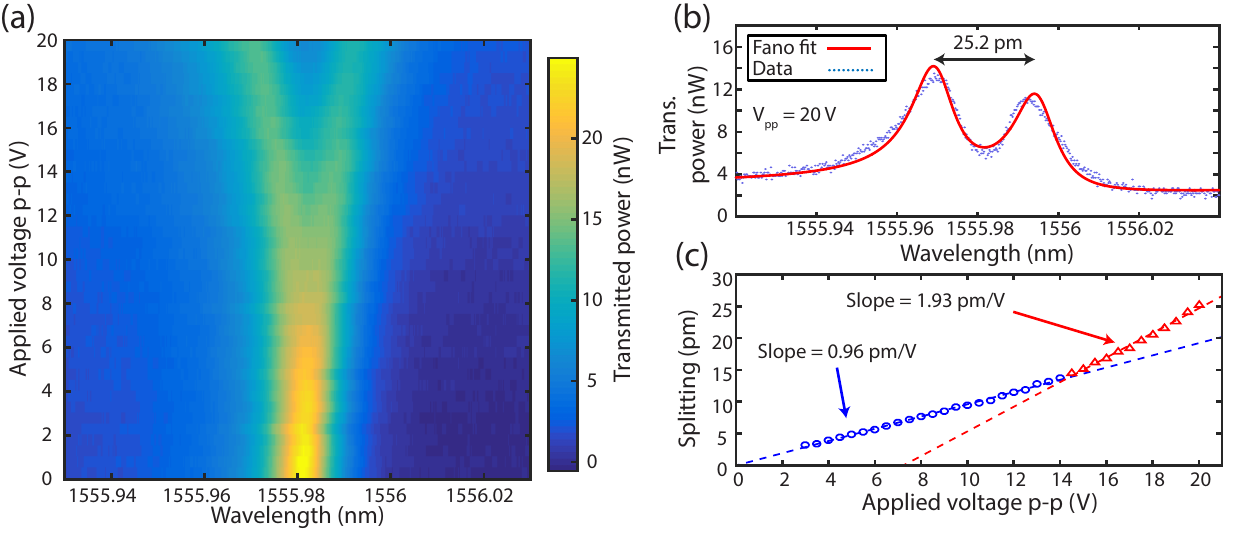}
\caption{(a) A density plot of showing the average transmitted power through the photonic crystal as a function of applied drive voltage and laser wavelength. An apparent splitting of the fundamental photonic crystal cavity mode is generated by applying the oscillating square-wave voltage.  (b) The fundamental cavity mode with an applied square-wave voltage of 20 V peak-to-peak.  The splitting of 25.2 pm is obtained by fitting a sum of two displaced lineshapes (red) of the undriven system to the data. (c) A piecewise linear fit to the splitting vs. applied voltage gives an EO modulation coefficient of 0.96 pm/V for lower voltages and 1.93 pm/V for larger voltages. }
\label{fig:electro-optic}
\end{center}
\end{figure}

To directly measure the EO modulation coefficient for the photonic crystals, a  function generator was used to apply a 1 MHz square-wave voltage to the electrodes.  Electrical contact was made by wire-bonding traces of a printed circuit board to aluminum contact pads on the SiLN chip.  The EO modulation is measured by observing the change in the optical spectrum as shown in Figure \ref{fig:electro-optic}.  A New Focus tunable laser is scanned while the cavity is being modulated and the transmitted light is sent to a detector whose bandwidth is set to be much smaller than the 1 MHz modulation frequency. Because the cavity frequency shift due to the applied voltage is much faster than the laser wavelength scan and detector bandwidth (1 $\mu$s compared to about 1 ms), the transmission spectrum shows an average of the two extreme mode positions.  This splitting increases with the magnitude of the applied voltage, as can be seen in the density plot in Figure \ref{fig:electro-optic} (a).

By fitting the doubly peaked lineshape (taking into account the slight Fano asymmetry apparent at $V=0$) we find the center wavelengths of the two peaks for each applied voltage. In this way we measure the splitting and extract the linear EO modulation coefficient.  An advantage of measuring the EO modulation coefficient in this way is that it makes the measurement insensitive to small offsets in the laser wavelength from scan to scan as well as robust to errors in calculating the optical powers.  Figure \ref{fig:electro-optic} (b) shows a single wavelength scan taken while applying a peak-to-peak voltage of 20 V.  Figure \ref{fig:electro-optic} (c) shows a piecewise linear fit to the peak splitting as a function of the applied voltage.  From this data the EO modulation coefficient is measured to be  0.96 pm/V up to around 14 V and 1.9 pm/V above 14 V.  This value is similar to measurements in Reference \citenum{Poberaj2012}, where they achieve modulation coefficient of 1.1 pm/V in LN rings with $Q$'s around 10,000.

From simulations of the electro-optic coefficient which treat the silicon layer as a perfectly insulating dielectric, we expect a modulation coefficient on the order of $3.6$ pm/V with a magnitude that is independent of the applied voltage. We believe the discrepancy between simulation and measurements are due to semiconductor field-effect physics in our device. The silicon that the devices are made from has a finite conductivity ($\approx$ 0.1 $\Omega^{-1}$ cm$^{-1}$) due to light p-doping. This means that some screening of the electric field in the LN will occur when the free carriers flow in the silicon in response to the RF field. This effect is expected to have a frequency dependence with an effective $RC$ time constant, and we measure a reduction of the screening at higher frequencies. Additionally, the applied field creates a region of depleted silicon without majority carriers.  As the applied voltage is increased the size of the depletion region also increases, until eventually the entire waveguide is effectively depleted of holes except for those localized near one of the boundaries.  At this point the screening effect is saturated and the EO modulation coefficient increases, as evidenced by the kink in Figure \ref{fig:electro-optic} (c).  In quasi-static semiconductor device simulations this saturation occurs at voltages on the order of 10 V. It should be noted that although we do see the effects of carrier depletion, we do not see evidence of an inversion layer, which would normally cause part of the depleted p-type silicon to become n-type and increase the electric field screening, causing a reduced slope in the curve of Figure \ref{fig:electro-optic}(c).  This is because inversion generates conduction electrons at a much slower rate than the modulation frequency so that the semiconductor cannot keep up with the modulated field. By moving to higher resistivity silicon and operating at higher frequencies, we expect to take advantage of the larger modulation coefficient in the operation of the quantum converter.

\section*{Conclusion}

We have demonstrated high-$Q$ electro-optically tunable photonic resonators on a chip. We have measured loss rates smaller than have been measured previously in lithium niobate microresonators and show that electro-optic coupling can be obtained in such a planar platform. In future work, these optical resonators can be combined with superconducting quantum circuits to enable quantum optical read-out and control of microwave circuits. This demonstration is a key step towards the implementation of an on-chip quantum electro-optic converter for emerging microwave-frequency quantum information systems.  In addition to quantum applications, we also expect that this platform will enable new high-sensitivity acousto-optic and electro-optic devices.


\section*{Acknowledgements}

JDW and PAA gratefully acknowledge support from a Stanford Graduate Fellowship. JAV was supported by Department of Physics Undergraduate Summer Research Fellowship. This work was supported by NSF ECCS-1509107 and the Stanford Terman Fellowship, as well as start-up funds from Stanford University. Part of this work was performed at the Stanford Nano Shared Facilities (SNSF) and the Stanford Nanofabrication Facility (SNF). ASN and JDW acknowledge valuable discussions with Martin Fejer and Carsten Langrock.

\section*{Author contributions statement}
JDW fabricated the devices, conceived and conducted the experiment, analyzed the data and wrote the manuscript.  JAV assisted in conducting the experiment.  PAA assisted in developing the fabrication.  CJS assisted in developing the experimental apparatus.  JTH assisted in the device fabrication and conducting the experiment. ASN conceived of the experiment and supervised the research.  All authors reviewed the manuscript. 

\section*{Additional information}

\textbf{Accession codes} \\
\textbf{Competing financial interests} \\
The authors declare that they have no competing financial interests.

\end{document}